\documentclass[12pt]{article}
\usepackage[english]{babel}
\usepackage{amsfonts}
\usepackage[dvips]{graphicx}

\begin{document}

\title{\bf Searching optimal shape in viscous flow: its dependence on Reynolds number.}

\date{October 2008}

\author{\bf Gianluca Argentini \\
\normalsize{[0,1]Bending - Italy}\\
\normalsize 01bending@gmail.com \\
\normalsize gianluca.argentini@gmail.com \\}

\maketitle

\begin{abstract}
In this work a simple problem on 2D optimal shape for body immersed in a viscous flow is analyzed. The body has geometrical constraints and its profile would be found in the class of cubics which satisfy those conditions. The optimal profile depends on the leading coefficient of these cubics and its relation with the Reynolds number of the system is found. The solution to the problem uses a method based on a suitable transformation rule for the cartesian reference.\\

\end{abstract}

We would study the problem of searching the optimal 2D shape or profile of a cartesian object immersed in a constant fluid viscous flow and subjected to some constraints on the boundary. This is a particular question on the general field of shape optimization for bodies immersed in flows (see \cite{pironneau}).\\

Let be $\{x,y\}$ a cartesian system, and $y = f(x)$ a function whose graph is the profile which we want to optimize. The function $f$ is subjected to the following constraints, arising from engineering requirements:\\

c1. $f(0) = 0$\\
\indent c2. $f(x_0) = y_0$\\
\indent c3. $f'(x_0) = 0$\\
\indent c4. $f'(0) \geq 0$\\
\indent c5. $f'(x) > 0$ $\hspace{0.1cm} \forall x \in \left(0,x_0\right)$\\

\noindent where $x_0$ and $y_0$ are positive numbers. The graph is immersed in a viscous flow having, at freestream zone, constant velocity ${\bf{V}}=(v_{\infty}, 0)$ parallel to the $x$-axis, with $v_{\infty} > 0$. What is, or which are, the functions $f$ that minimize the pressure drop along its profile, that is, if $p = p(x,y)$ is the static pressure along the graph, what is the curve for which the difference $\Delta p = p(0,0) - p(x_0,y_0)$ is minimum? The question is related to drag and lift optimization (see \cite{dede}).\\

We consider the cubics $f(x) = ax^3 + bx^2 + cx +d$. Using c1, c2 and c3, we find

\begin{eqnarray}
	\nonumber b &=& - \frac{y_0 +2ax_0^3}{x_0^2}\\
	c &=& \frac{2y_0+ax_0^3}{x_0}\\
	\nonumber d &=& 0
\end{eqnarray}

\noindent The derivative of the general cubic is $f'(x) = 3ax^2 + 2bx +c$ so that $f'(0) = c$. From c4, we have the following condition on parameter $a$:

\begin{equation}\label{ac4Condition}
	a \geq -2\frac{y_0}{x_0^3}
\end{equation}
	
\noindent while a few algebra, imposing the condition $f'(t)=0$ if and only if $t \leq 0$ or $t \geq x_0$, shows that c5 is satisfied if

\begin{equation}\label{ac5Condition}
	a \leq \frac{y_0}{x_0^3}
\end{equation}

\noindent Then the search of the optimal profile is defined on the class of cubics

\begin{equation}\label{class}
	C = \left\{ ax^3 - \frac{y_0 +2ax_0^3}{x_0^2}x^2 + \frac{2y_0+ax_0^3}{x_0}x \hspace{0.1cm} : \hspace{0.1cm} -2\frac{y_0}{x_0^3} \leq a \leq \frac{y_0}{x_0^3} \right\}
\end{equation}

\begin{figure}[h!]\label{cubics}
	\begin{center}
	\includegraphics[width=10cm]{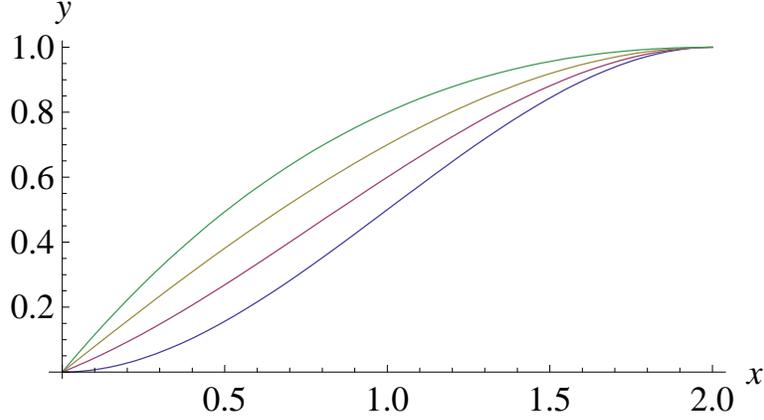}
	\caption{\it Some cubics in the case $x_0=2$, $y_0=1$: from $a = -0.25$ (blue curve) to $a = 0.125$ (green). }
	\end{center}
\end{figure}

We use Newton's theory on fluid velocity distribution on the profile or surface of a body immersed in a flow with constant freestream speed $v_{\infty}$ (\cite{edwards}). If $\alpha = \alpha(x)$ is the angle between the direction of the flow and the tangent to the curve, the velocity in a point of the profile has two components, the tangential one  with module $v_{\infty} cos\alpha$, and the normal one with module $v_{\infty} sin\alpha$. The latter gives the amount for the body resistence (see \cite{lachand}). But we would analyze the flow on the upper neighbour of the profile, where fluid has a tangent velocity field given by

\begin{equation}\label{fluidField1}
	{\bf v} = (v_{\infty}cos^2\alpha, v_{\infty}sin\alpha \hspace{0.1cm} cos\alpha)
\end{equation}

\noindent From usual trigonometrical formulas the following identity holds

\begin{equation}\label{trig}
	cos^2\alpha(x) = \frac{1}{1+f'(x)^2}
\end{equation}

\noindent so that we have

\begin{equation}\label{fluidField2}
	{\bf v} = (v_1, v_2)= \left(v_{\infty}\frac{1}{1+f'^2}, v_{\infty}\frac{f'}{1+f'^2}\right)
\end{equation}

Let $\mu$ the dynamic viscosity, $\rho$ the density and $p=p(x,f(x))$ the static pressure of the fluid on the body profile. We could write the Navier-Stokes equations for this flow in the upper neighbour of the profile, with (\ref{fluidField2}) and $p(0,0) = p(x_0,y_0)$ as boundary condition:

\begin{eqnarray}
\left\{
\begin{array}{ll}
	\nonumber \rho \hspace{0.1cm} \nabla{\bf v} \hspace{0.1cm} {\bf v} = - \nabla p + \mu \Delta{\bf v}\\
	{\bf v}(x,f(x)) = v_{\infty} \left( \frac{1}{1+f'^2}, \frac{f'}{1+f'^2} \right)\\
	\nonumber p(0,0) = p(x_0,y_0)
\end{array}
\right.
\end{eqnarray}

\noindent We try to simplify the resolution of previous system by the following transformation rule on coordinates system:

\begin{equation}\label{transformationRule}
\left\{
\begin{array}{ll}
	Y = y - f(x)\\
	X = f(x)
\end{array}
\right.
\end{equation}

\noindent It is important to note that, from c5 condition and usual notions of real analysis, the function $f$ is invertible on the interval $(0, x_0)$ with inverse $g$ differentiable: $d_X g = \frac{1}{d_x f}$.\\
\noindent In the new coordinates system $\{X,Y\}$, the curve $y = f(x)$ has the simple equation $Y = 0$, hence it becomes the $X$-axis itself. 

\begin{figure}[h!]\label{newXY}
	\begin{center}
	\includegraphics[width=6cm]{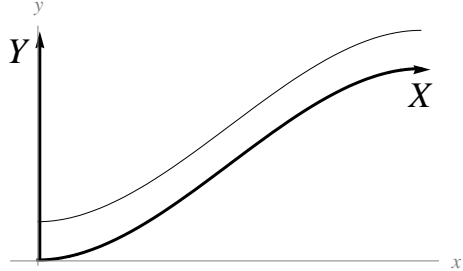}
	\caption{\it The new cartesian system $XY$, from the old $xy$ point of view, in the case $a = -0.25$. If $X=0$ and $y=0$, then $x=0$ and $Y=0$ so that the new origin is the same as the old. For $Y=0$ we have $y=f(x)$, so that the new $X$-axis is the curve $f$ in the old system. If $X=0$, then $f(x)=0$ and $Y=y$, so that the new $Y$-axis is the old $y$-axis in the $xy$ system. The thin line has equation of type $Y=cost$.}
	\end{center}
\end{figure}

Then, the ($X$,$Y$)-flow near the curve is parallel to $X$, that is the velocity field ${\bf U}$ in the system $\{X,Y\}$ has only the $X$-component: ${\bf U} = (U_1, 0)$. If $\sigma = \rho(g(X),Y+X)$, $\nu = \mu(g(X),Y+X)$ and $P = p(g(X),Y+X)$ are the representation of the scalar functions $\rho$, $\mu$, $p$ in the $\{X,Y\}$ system, the flow is described by the simple Navier-Stokes equation

\begin{equation}\label{newNSeq}
	\sigma U_1 \frac{\partial U_1}{\partial X} = - \frac{\partial P}{\partial X} + \nu \frac{\partial^2 U_1}{\partial X^2}
\end{equation}

\noindent The boundary conditions are now $U(X,0) = v_1(g(X),Y+X)$ and $P(0,0) = P(y_0,0)$.\\

We can write the expression of $U_1$,

\begin{equation}\label{U1}
	U_1 = U_1(X,Y) = v_1(g(X),Y+X) = v_{\infty}\frac{g'^2}{1+g'^2}
\end{equation}

\noindent where $g' = d_X g$, and the expression of its first derivative:

\begin{equation}\label{partialU1}
	\frac{\partial U_1}{\partial X} = v_{\infty}\frac{2g'g''(1+g'^2)-2g'^3g''}{(1+g'^2)^2} = 2v_{\infty}\frac{g'g''}{(1+g'^2)^2}
\end{equation}

\noindent Note that $U_1$ doesn't depend explicitly on $Y$. For $X=f(x_0)=y_0$ and $Y=0$, the value of $U_1$ is defined by continuity extension, because $f'(x_0)=0$ and consequently $d_Xg(y_0)=\infty$. From (\ref{U1})

\begin{equation}
	U_1(y_0,0) = \lim_{X \rightarrow y_0} v_{\infty}\frac{g'^2}{1+g'^2} = v_{\infty}\lim_{X \rightarrow y_0} \frac{1}{\frac{1}{g'^2}+1} = v_{\infty}
\end{equation}

\noindent In the case $f'(0)=0$, at the same manner the definition of $U_1$ can be extended at $(0,0)$ by $U_1(0,0)=v_{\infty}$.\\
\noindent Now we integrate the two members of (\ref{newNSeq}) respect to the $X$ variable from $0$ to $f(x_0)=y_0$:

\begin{equation}\label{tempOptEquation1}
	\frac{1}{2} \sigma \left[U_1^2\right]_0^{y_0} = - \left[P\right]_0^{y_0} + \nu \left[\partial_X U_1\right]_0^{y_0}
\end{equation}

\noindent For our purpose of optimization, we impose $- \left[P\right]_0^{y_0} = P(0,0)-P(y_0,0) = p(0,0) - p(x_0,y_0) = 0$.\\
\noindent At first consider the simple case of inviscid flow: $\nu = 0$. Previous equation becomes

\begin{equation}
	\frac{1}{2} \sigma \left[U_1^2\right]_0^{y_0} = 0
\end{equation}

\noindent therefore it must be $U_1(0,0) = U_1(y_0,0)$. Using previous identities, this equation can be written as

\begin{equation}
		\frac{1}{2} \sigma v_{\infty}^2 \left[ 1 - \frac{1}{(1+f'(0)^2)^2} \right] = 0
\end{equation}

\noindent Then it must be $f'(0)=0$, that is $c = \frac{2y_0+ax_0^3}{x_0} = 0$. Therefore, in the case of inviscid flow, the optimization problem is solved by the cubic, which belongs to the class $C$, with

\begin{equation}\label{aSolutionInviscid}
	a = - 2\frac{y_0}{x_0^3}
\end{equation}

\noindent The cartesian equation is $y = - 2\frac{y_0}{x_0^3} x^3 + 3\frac{y_0}{x_0^2} x^2$.\\

Now we consider the viscous case. The condition to impose in equation (\ref{tempOptEquation1}) is always $\left[P\right]_0^{y_0} = 0$, therefore

\begin{equation}\label{tempOptEquation2}
	\frac{1}{2} \sigma \left[U_1^2\right]_0^{y_0} = \nu \left[\partial_X U_1\right]_0^{y_0}
\end{equation}

\noindent The first member is

\begin{equation}\label{firstMember}
	\frac{1}{2} \sigma v_{\infty}^2 \left[ 1 - \frac{1}{(1+f'(0)^2)^2} \right]
\end{equation}

\noindent For expliciting the second member, from (\ref{partialU1}) we have to find $g''=g_{XX}$. Apply usual differentiation rules:

\begin{eqnarray}
	g_{XX}(X) = d_X g_X(X) = d_X \frac{1}{d_x f(g(X))} = \\
	\nonumber = d_x \frac{1}{f'}d_X x = -\frac{f''}{f'^2}g_X = - f''(g(X)) g_X^3(X)
\end{eqnarray}

\noindent Therefore we can write

\begin{eqnarray}
	\partial_X U_1(y_0) = - \lim_{X \rightarrow y_0}2 v_{\infty} f''(g(X))\frac{g_X^4}{(1+g_X^2)^2} = - 2 v_{\infty} f''(x_0)\\
	\noindent \partial_X U_1(0) = - \lim_{X \rightarrow 0}2 v_{\infty} f''(g(X))\frac{g_X^4}{(1+g_X^2)^2} = - 2 v_{\infty} \frac{f''(0)}{(1+f'(0)^2)^2}
\end{eqnarray}

\noindent For a cubic of the class $C$ the following identities hold: $f'(0) = \frac{2y_0+ax_0^3}{x_0}$, $f''(0) = -2\frac{y_0+2ax_0^3}{x_0^2}$ and $f''(x_0) = 2\frac{-y_0+ax_0^3}{x_0^2}$. Equation (\ref{tempOptEquation2}) can be written as

\begin{equation}
	\sigma v_{\infty} \left[ 1 - \frac{1}{(1+f'(0)^2)^2} \right] = 4\nu \left[ - f''(x_0) + \frac{f''(0)}{(1+f'(0)^2)^2} \right]
\end{equation}

\noindent or

\begin{equation}
  \sigma v_{\infty} \left[(1+f'(0)^2)^2 -1 \right] = 4\nu \left[f''(0) - (1+f'(0)^2)^2 f''(x_0) \right]
\end{equation}

\noindent It is an algebraic equation of the 5th order in the parameter $a$. But we are interested in physical situations where viscosity is small, that is when the parameter $a$ has values near the previous computed quantity (\ref{aSolutionInviscid}): then, in this situation, $f'(0)$ is small, $f''(0)=2b=6\frac{y_0}{x_0^2}$ and $f''(x_0)=-6\frac{y_0}{x_0^2}=-f''(0)$. We can consider the following Taylor expansion of the quantity $(1+f'(0)^2)^2$:

\begin{equation}
	(1+f'(0)^2)^2 \approx 1 + 2f'^2(0) = 1 + 2\frac{(2y_0+ax_0^3)^2}{x_0^2}
\end{equation}

\noindent Then the algebraic equation in the parameter $a$ can be simplified in the following 2nd order one:

\begin{equation}
	\left(1 + 2\frac{(2y_0+ax_0^3)^2}{x_0^2} \right) \left( 4 \nu f''(0) -\sigma v_{\infty} \right) + \left( 4 \nu f''(0) + \sigma v_{\infty} \right) = 0
\end{equation}

\noindent The admissible solution to this equations is

\begin{equation}\label{aSolutionViscous}
	a = - 2 \frac{y_0}{x_0^3} + \frac{1}{x_0^2}\sqrt{\frac{6 \nu y_0}{| 24 \nu y_0-\sigma v_{\infty}x_0^2 |}}
\end{equation}

\noindent Note that, in the case $\nu = 0$, the solution has the expression (\ref{aSolutionInviscid}), as expected. In viscous case, on the contrary as inviscid one, the optimal profile depends on the values of density $\rho$, viscosity $\mu$, and freestream speed $v_{\infty}$.\\
\noindent We can see that the main effect of viscosity is its influence on $f'(0)$, that is on the angle of attack between flow and profile. In fact, $a$ is an increasing function of $\mu$, therefore $f'(0) = c = \frac{2y_0+ax_0^3}{x_0}$ is an increasing function too.

\begin{figure}[h!]\label{aParameterVisc}
	\begin{center}
	\includegraphics[width=8cm]{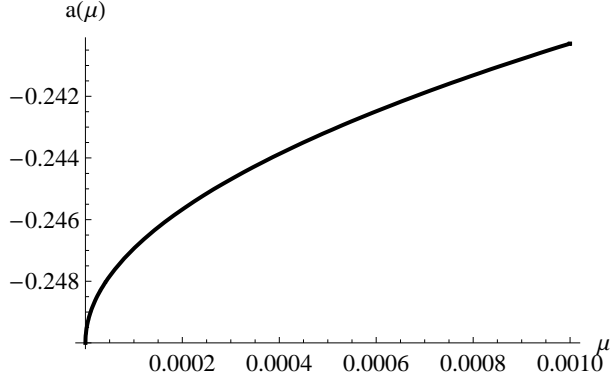}
	\caption{\it Parameter $a$ as function of $\mu$ in the case $x_0=2$, $y_0=1$, $\sigma=0.001$ and $v_{\infty}=1000$ $(CGS \hspace{0.2 cm}units)$. }
	\end{center}
\end{figure}

\noindent If we consider a flow with a constant value of $\mu$, it is interesting to evaluate $a$ as function of the freestream speed $v_{\infty}$. From (\ref{aSolutionViscous}) follows that $a$ is a decreasing function of $v_{\infty}$ and

\begin{equation}
	\lim_{v_{\infty} \rightarrow \infty} a(v_{\infty}) = - 2\frac{y_0}{x_0^3}
\end{equation}

\noindent therefore the increasing of flow speed is equivalent to a vanishing of viscosity.\\

Now multiply by $y_0$ and divide by $\nu x_0^2$ both numerator and denominator of the radicand in (\ref{aSolutionViscous}). Introducing the label $z_0 = \frac{y_0}{x_0}$ and the {\it Reynolds number}

\begin{equation}\label{reynolds}
	{\mathbb R}e = \frac{\sigma v_{\infty} y_0}{\nu}
\end{equation}

\noindent (recall that $\sigma$ is the density and $\nu$ the dynamic viscosity) where $y_0$ is the characteristic length of this geometrical system (see \cite{pironneau}), the parameter $a$ can be written in the form

\begin{equation}\label{aSolutionViscousRe}
	a = \frac{1}{x_0^2} \left( -2z_0 + \sqrt{\frac{6z_0^2}{|24z_0^2 - {\mathbb R}e |}} \right)
\end{equation}

\noindent This expression separates the dependence of $a$ on geometrical ($x_0$, $z_0$) and physical parameters (${\mathbb R}e$). As expected, for high Reynolds number the influence of viscosity vanishes and the optimal profile tends to the shape of the case $\mu = 0$:

\begin{equation}
	\lim_{{\mathbb R}e \rightarrow \infty} a({\mathbb R}e) = \frac{1}{x_0^2}(-2z_0) = - 2 \frac{y_0}{x_0^3}
\end{equation}

\begin{figure}[h!]\label{aParameterRe}
	\begin{center}
	\includegraphics[width=8cm]{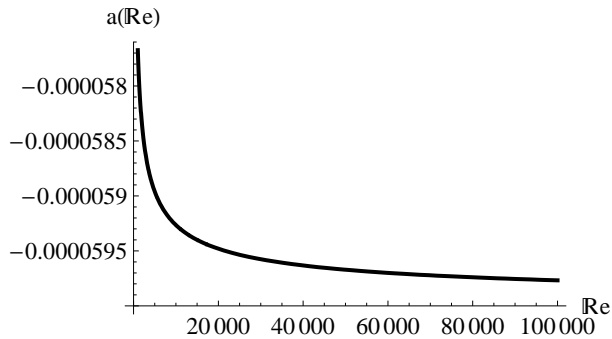}
	\caption{\it Parameter $a$ as function of ${\mathbb R}e$ in the case $x_0=100$, $z_0=0.3$ $(CGS \hspace{0.2 cm}units)$.}
	\end{center}
\end{figure}

\begin{figure}[h!]\label{OptimalProfiles}
	\begin{center}
	\includegraphics[width=12cm]{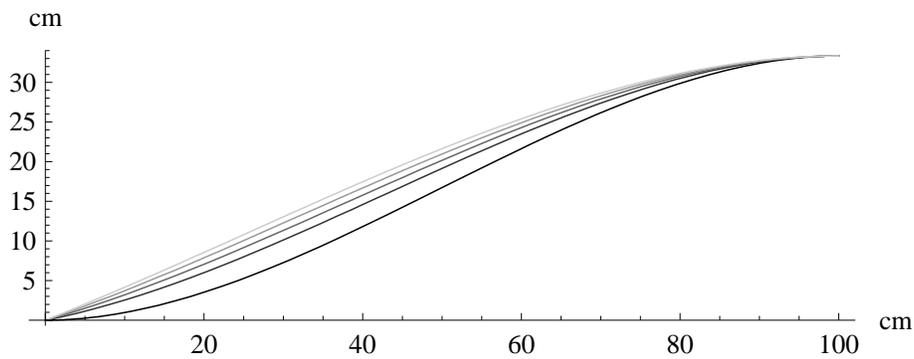}
	\caption{\it Optimal profile as function of $\mu \in [0.0, 0.001]$ in the case $x_0=100$, $z_0=0.3$ $(CGS \hspace{0.2 cm}units)$; decreasing grey color levels are associated to increasing values of viscosity.}
	\end{center}
\end{figure}

\begin{figure}[h!]
	\includegraphics[width=2cm]{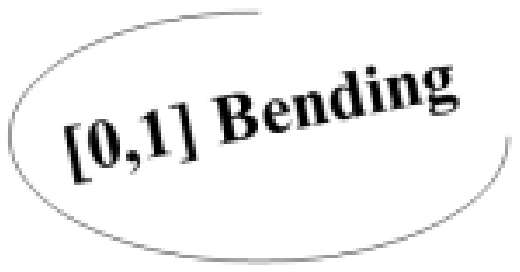}
\end{figure}
\noindent \tiny {\bf Gianluca Argentini}, mathematician, works on the field of\\
fluid dynamics and optimization of shapes for bodies\\
moving inside fluid flows. He has found {\bf [0,1]Bending},\\
a Design Studio in Italy for scientific and industrial applications.

\end{document}